\documentclass{mem}
\usepackage{natbib}\usepackage{txfonts}\usepackage{balance}
\usepackage{graphicx}
\usepackage[a4paper,breaklinks,dvipdfm]{hyperref}
\idline{75}{1}
\begin{document}
\def\teff{$T\rm_{eff }$}
\def\meth{CH$_4$}
\def\micron{$\mu$m}
\def\ldl{$\lambda/\Delta\lambda$}
\def\kms{$\mathrm {km s}^{-1}$}

\title{
Luhman 16AB: A Remarkable, Variable L/T Transition Binary 2 pc from the Sun
}

   \subtitle{}

\author{
A.\ J.\ Burgasser$^1$, 
J.\ Faherty$^2$, 
Y.\ Beletsky$^3$, 
P.\ Plavchan$^4$,  
M.\ Gillon$^5$, 
J.\ Radigan$^6$,
E. Jehin$^5$, 
L. Delrez$^5$,  
C. Opitom$^5$, 
N.\ Morrell$^3$, 
R.\ Osten$^6$, 
R.\ Street$^7$,
C.\ Melis$^1$, 
A.\ Triaud$^8$, 
\and R.\ Simcoe$^8$          }

  \offprints{A.\ J.\ Burgasser}

\institute{
$^1$ UC San Diego, 9500 Gilman Drive, Mail Code 0424, La Jolla, CA 92093, USA \\
$^2$ Universidad de Chile, Camino El Observatorio \#1515, Las Condes, Chile \\
$^3$ Las Campanas Observatory, Carnegie Observatories, Casilla 601, La Serena, Chile \\
$^4$ Caltech/IPAC, 770 S Wilson Ave Pasadena, CA 91125, USA \\
$^5$ Universit{\'{e}} de Li{\`{e}}ge, All{\'{e}}e du 6 ao\^{u}t 17, Sart Tilman, Li\`{e}ge 1, Belgium \\
$^6$ Space Telescope Science Institute, 3700 San Martin Dr., Baltimore, MD 21218, USA \\
$^7$ Las Cumbres Observatory, 6740 Cortona Drive Suite 102, Goleta, CA  93117, USA \\
$^8$ MIT, 77 Massachusetts Avenue, Cambridge, MA 02139, USA \\
\email{aburgasser@ucsd.edu}
}

\authorrunning{Burgasser et al.}

\titlerunning{Luhman 16AB}

\abstract{
Luhman (2013) has reported the discovery of a brown dwarf binary system only 2.01$\pm$0.15~pc from the Sun.  The binary is well-resolved with a projected separation of 1$\farcs$5, and spectroscopic observations have identified the components as late-L and early-T dwarfs.
The system exhibits several remarkable traits, including a ``flux reversal'', where the T dwarf is brighter over 0.9--1.3~$\mu$m but fainter at other wavelengths; and significant ($\sim$10\%) short-period ($\sim$4.9~hr) photometric variability with a complex light curve.  These observations suggest spatial variations in condensate cloud structure, which is known to evolve substantially across the L dwarf/T dwarf transition. Here we report preliminary results from a multi-site monitoring campaign aimed at probing the spectral and temporal properties of this source.  Focusing on our spectroscopic observations, we report the first detections of NIR spectral variability, present detailed analysis of K~I lines that confirm differences in condensate opacity between the components; and preliminary determinations of radial and rotational velocities based on high-resolution NIR spectroscopy. 
}
\maketitle{}

\section{Introduction}

The transition between the L dwarf and T dwarf spectral classes has emerged as one of the outstanding problems in brown dwarf astrophysics.
Spectroscopically, this transition is defined by the appearance of {\meth} absorption at near-infrared (NIR) wavelengths (e.g., \citealt{2006ApJ...637.1067B}) and a reduction in condensate cloud opacity (e.g., \citealt{2002ApJ...568..335M}), both leading to large changes in NIR spectral energy distributions (SEDs). 
Among field brown dwarfs, this transition takes place 
over a surprisingly narrow range of effective temperatures ($\Delta${\teff} $\approx$ 100--200~K) and luminosities ($\Delta{M_{bol}}$ $\approx$ 0.1--0.3~dex; \citealt{2004AJ....127.3516G,2009ApJ...702..154S}).
Even more remarkable is the existence of so-called ``flip'' binaries, whose components straddle the L/T transition and in which the later-type secondary can be brighter at 1~{\micron} than the primary
\citep{2006ApJS..166..585B,2006ApJ...647.1393L}.
The most likely origin for these trends is a universal, ``rapid'' and possibly heterogeneous depletion of photospheric condensates at the L/T transition, the mechanism of which remains unclear to this day \citep{2001ApJ...556..872A,2002ApJ...571L.151B,2004AJ....127.3553K,2008ApJ...689.1327S}.

The recently-discovered binary brown dwarf WISE~J104915.57$-$531906.1AB (hereafter Luhman~16AB; \citealt{2013ApJ...767L...1L}) has emerged as a key laboratory for studying the L/T transition.
Its components straddle the transition, and it is a flip binary \citep{2013ApJ...770..124K,2013arXiv1303.7283B}.  Its proximity (2.02$\pm$0.15~pc) implies its components are both easily resolved (1$\farcs$5) and bright, permitting detailed investigation of their atmospheres; yet the system is compact enough for orbit determination in a 20--30~yr timeframe. Finally, this source has been shown to be variable \citep{2013arXiv1304.0481G} with a peak-to-peak amplitude of $\sim$10\% and period of 4.87$\pm$0.01~hr.  

\section{The Monitoring Campaign}

\begin{table*}[th]
\caption{Observations Conducted as Part of the Luhman~16AB Monitoring Campaign}
\label{obs}
\begin{center}
\begin{tabular}{llll}
\hline
\\
Instrument & Program & Leads  \\
\hline
ESO/TRAPPIST & Combined red optical monitoring & Gillon, Triaud  \\
DuPont/RETROCAM & Resolved NIR monitoring & Faherty, Morell, Radigan  \\
NTT/SOFI & Combined NIR monitoring & Radigan  \\
CTIO/ANDICAM & Resolved optical \& NIR monitoring & Faherty, Radigan  \\
Magellan/FIRE & Resolved moderate resolution NIR spectroscopy & Faherty, Burgasser  \\
Magellan/MagE & Resolved moderate resolution optical spectroscopy & Beletsky, Faherty  \\
Magellan/MIKE & Resolved high resolution optical spectroscopy & Beletsky  \\
IRTF/CSHELL & Resolved high resolution NIR spectroscopy & Plavchan, Burgasser \\
IRTF/SpeX & Resolved low resolution NIR spectral monitoring & Burgasser  \\
ATCA & Resolved radio monitoring & Osten, Melis, Radigan \\
Las Cumbres Network & Resolved optical monitoring & Street & \\
\hline
\end{tabular}
\end{center}
\end{table*}

\begin{figure}[t!]
\centering
\resizebox{0.9\hsize}{!}{
\includegraphics[angle=90]{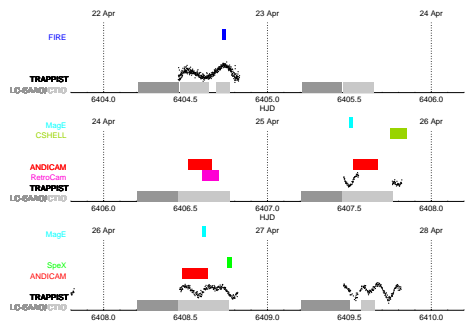}}
\caption{\footnotesize
Scheduling blocks of observations over 22-28 April 2013 (UT). 
ESO/TRAPPIST observations are indicated by the plotted lightcurve.}
\label{fig:schedule}
\end{figure}

Given its unique characteristics and importance for understanding the L/T transition, we coordinated a week-long monitoring campaign of Luhman~16AB using telescopes in Chile, Australia, Hawaii and South Africa. Our aim was to measure the components of this system panchromatically (radio, optical and infrared), spectroscopically (optical and near-infrared, low to high resolution), and temporally.
The observations conducted are summarized in Table~\ref{obs} and illustrated in Figure~\ref{fig:schedule}. Throughout this period, ESO/TRAPPIST \citep{2011Msngr.145....2J} was used to anchor the observations to a common light curve. These data show that Luhman~16AB has persisted in its strong variability.
All of our spectroscopic observations coincided with some period of TRAPPIST monitoring; the entire period is also blanketed by monitoring observations from the Las Cumbres Observatory Global Telescope Network (LCOGT; \citealt{2012IAUS..285..408S}).

Here we report preliminary results from the spectroscopic components of our campaign.

\subsection{Resolved FIRE Spectroscopy}

\begin{figure*}[th]
\resizebox{\hsize}{!}{
\includegraphics[clip=true]{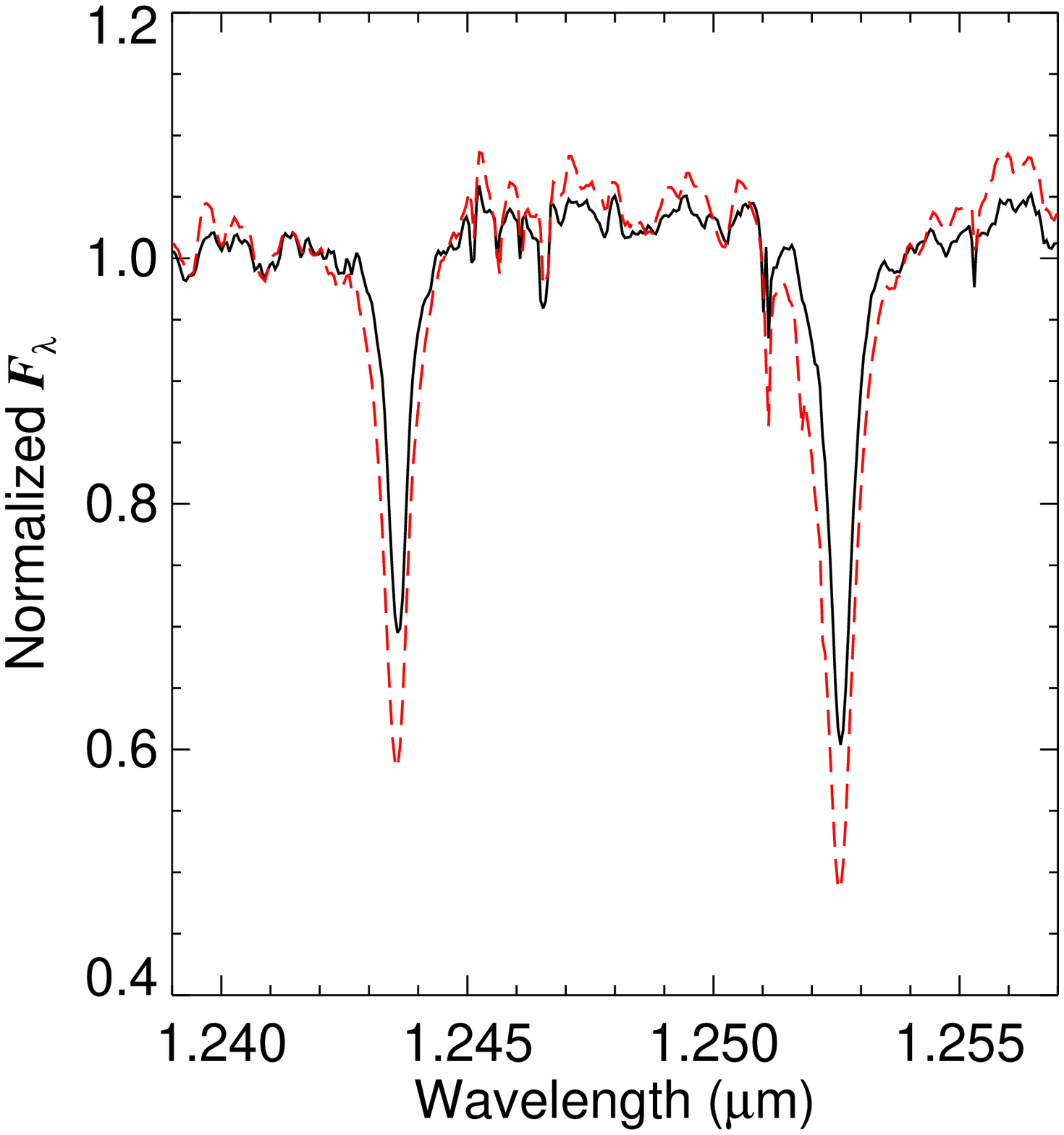}
\includegraphics[clip=true]{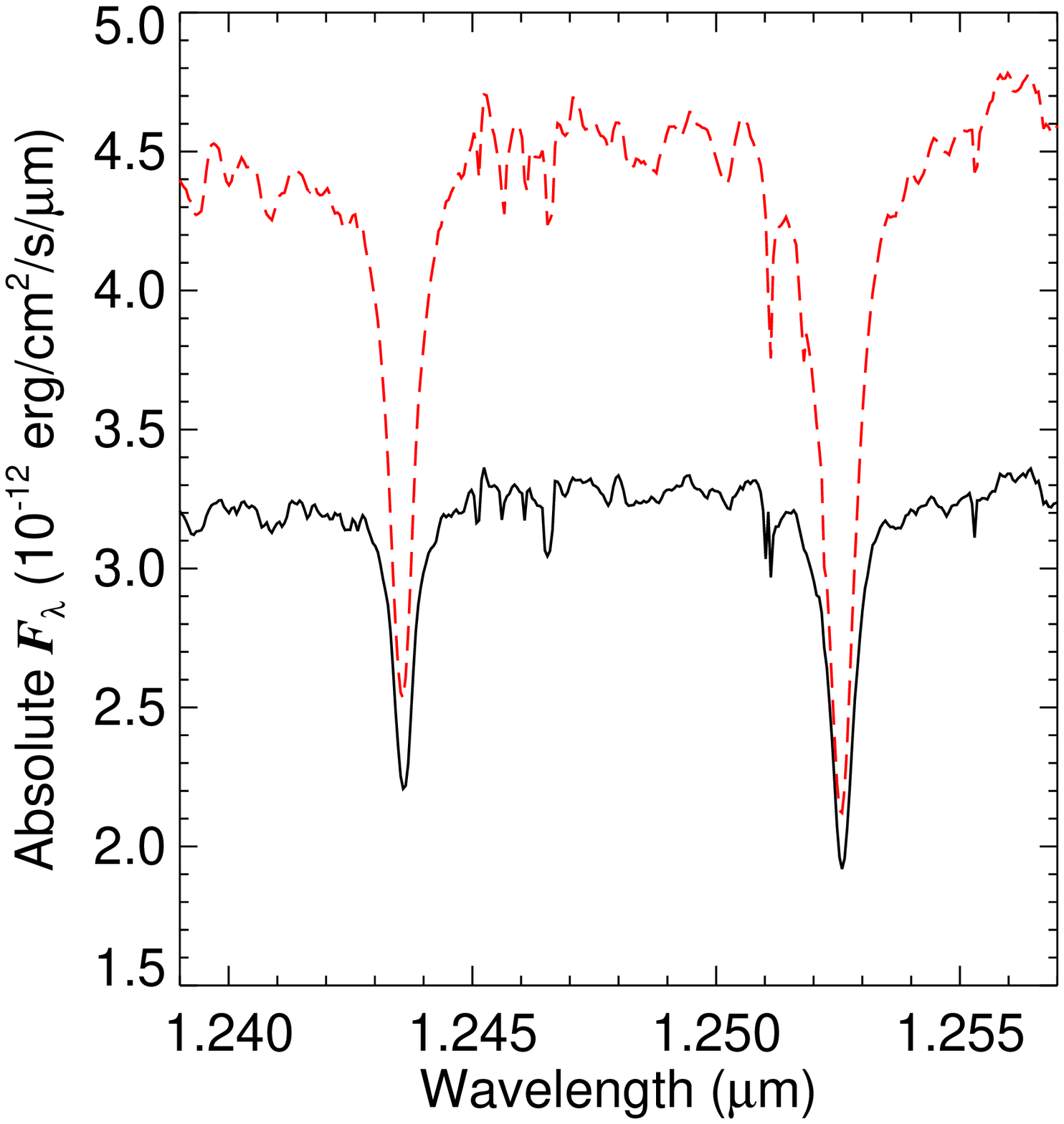}
\includegraphics[clip=true]{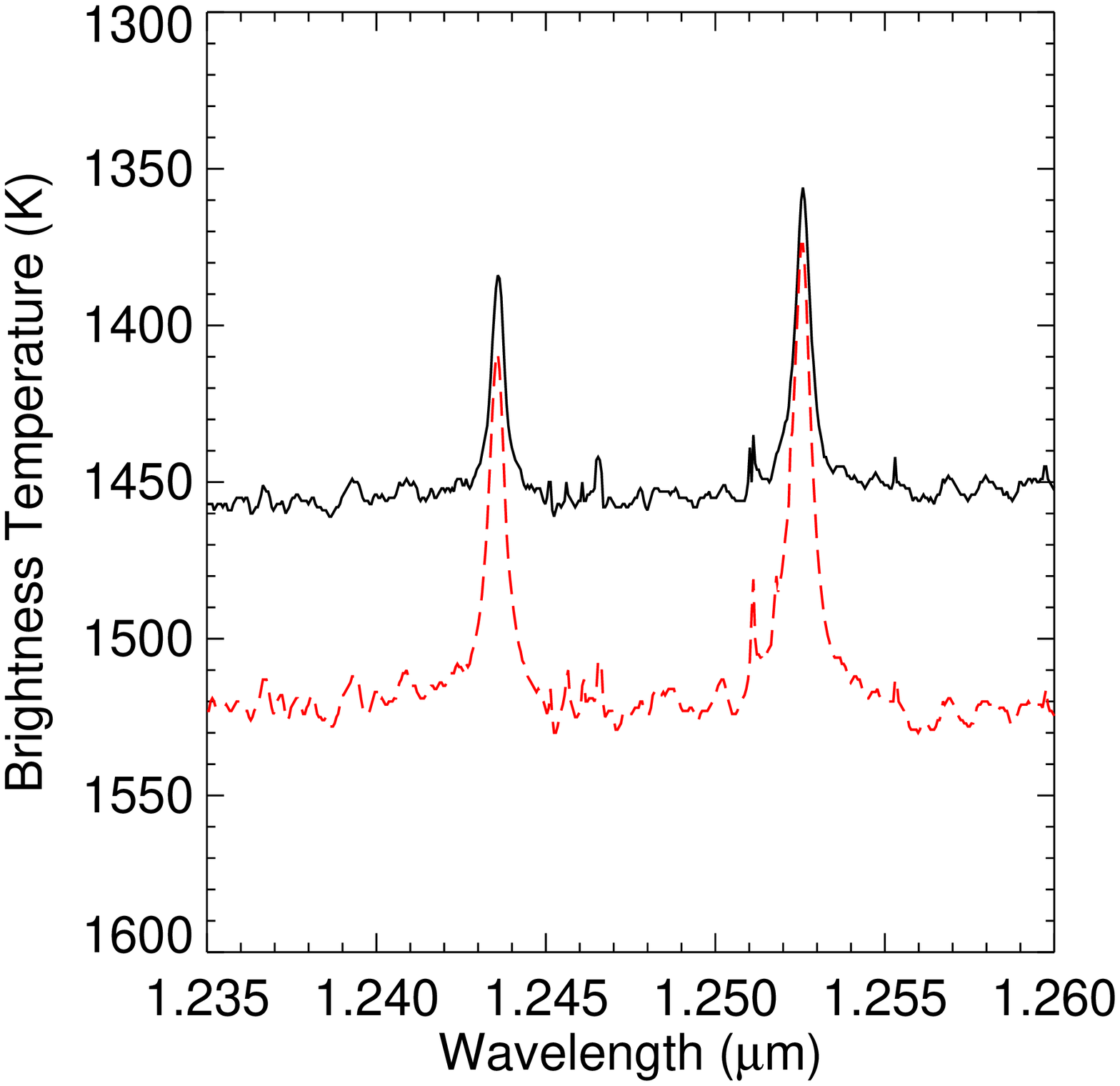}}
\caption{\footnotesize
The 1.25~{\micron} K I doublet in FIRE data for Luhman~16A (solid black lines) and B (dashed red lines).  (Left): Normalizing at the local continuum, it appears that Luhman~16B has deeper, broader lines, suggesting differences in abundances or $v\sin{i}$. (Middle): However, when scaled to absolute fluxes, it is clear that it is the continuum, not the lines, that varies between these sources. (Right): We can equate the absolute fluxes to brightness temperatures assuming radii R = 1~R$_{Jup}$.  The $\sim$70~K offset in the continuum between Luhman~16A and B (note inverse scale) can be attributed to reduced condensate opacity in the latter.}
\label{fig:k1}
\end{figure*}

Early NIR spectroscopy of Luhman~16AB were limited to low-resolution data (e.g., \citealt{2013arXiv1303.7283B}).  In contrast, the Folded-port Infrared Echellette (FIRE; \citealt{2010SPIE.7735E..38S}) on the Magellan 6.5m Baade Telescope provides 
moderate-resolution ({\ldl} $\approx$ 8000) spectroscopy spanning the 0.8--2.4~{\micron} band. We obtained resolved spectra on 22 April 2013 (UT), and several epochs thereafter, achieving S/N $\approx$ 500 in 200~s integration per source.  The spectra reveal remarkable detail in atomic and molecular features, including detection of FeH and CH$_4$ absorption at $H$-band in both components, and clear detections of several alkali features at $\lambda < 1.3$~{\micron}. Focusing on the 1.25~{\micron} K~I doublet (Figure~\ref{fig:k1}), these lines are sufficiently strong that we can rule out either component having a low surface gravity, as would be expected if the system was a member of the $\sim$40~Myr Argus association \citep{2013arXiv1303.5345M}.

Comparing the line shapes between sources, by normalizing at the continuum it appears that Luhman~16B has stronger, broader lines, consistent with previously detected upticks in K~I equivalent widths across the L/T transition \citep{2002ApJ...564..421B}.  However, when the spectra are normalized to their absolute magnitudes, we find that the lines are simply nested, with Luhman~16B having a brighter continuum. By associating spectral fluxes with brightness temperatures (T$_{br}$), we infer a $\sim$70~K difference in the photospheric temperature of these two dwarfs in the 1.25~{\micron} continuum, with Luhman~16B being the hotter source.  Importantly, this T$_{br}$ offset is {\em not present} in regions where molecular gas opacity dominates the continuum (e.g., around the 1.175~{\micron} K~I doublet).  As the continuum around 1.25~{\micron} in L dwarfs is dominated by condensate grain scattering opacity (e.g., \citealt{2001ApJ...556..872A}), we attribute the T$_{br}$ offset here to reduced condensate opacity in Luhman~16B.  Further analysis of these data will be presented in Faherty et al.\ (in prep.).

\subsection{Spectral Monitoring with SpeX}

\begin{figure}[th]
\resizebox{\hsize}{!}{\includegraphics[clip=true]{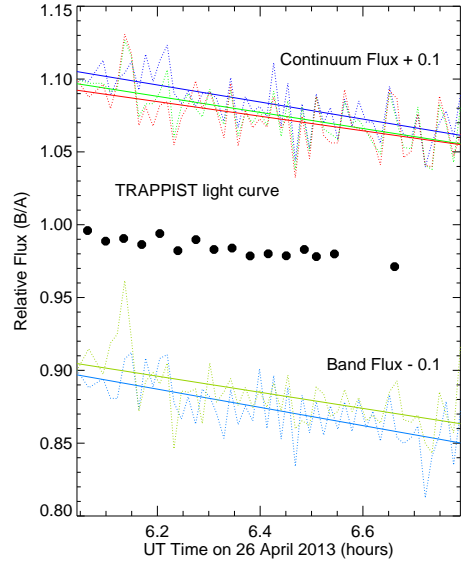}}
\caption{\footnotesize
Variations in relative flux (B/A) in several spectral regions sampling continuum and absorption band regions.  The uniform declines are consistent with TRAPPIST combined-light data.}
\label{fig:monitor}
\end{figure}

Luhman~16AB was spectroscopically monitored with SpeX \citep{2003PASP..115..362R} on the 3m NASA Infrared Telescope Facility (IRTF) on 26 April 2013 (UT), using that instrument's low-resolution ({\ldl} $\approx$ 120) prism-dispersed mode.  
The system was observed for 45~min with the slit aligned along the binary axis for simultaneous spectroscopy.  As seeing was comparable to the binary separation, the data were forward-modeled using a 10-parameter profile model, then calibrated using standard techniques.   These new observations confirm the ``flip'' reported in \citet{2013arXiv1303.7283B}, apparent even in the raw data.
Slit losses and differential color refraction\footnote{An illustration of how color refraction creates coordinated variability can be seen at \url{http://www.youtube.com/watch?v=DIJx0flF6uc}.} limit our analysis to {\em relative} spectral fluxes between the components (B/A).  Figure~\ref{fig:monitor} shows that these relative fluxes decline over the observing period, coincident with the TRAPPIST lightcurve; this is consistent with a dimming of Luhman~16B.  The decline notably occurs for regions sampling both low and high gas opacity.  Simllar achromatic flux variations have been reported in other variable L/T dwarfs \citep{2013ApJ...768..121A}, suggesting a common origin.

\subsection{High-resolution Spectroscopy with CSHELL}

\begin{figure*}[th]
\resizebox{0.95\hsize}{!}{
\includegraphics[clip=true]{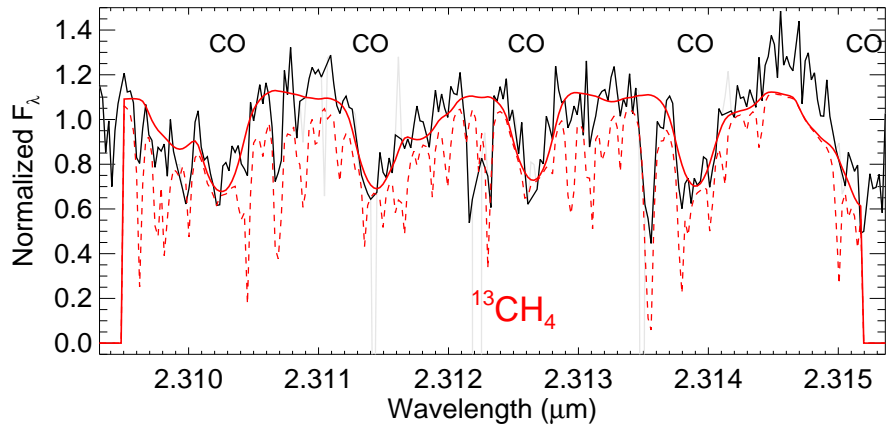}}
\caption{\footnotesize
Preliminary analysis of IRTF/CSHELL data, comparing the observed combined-light spectrum (black line) to a spectral model (solid red line) with $v_{rad}$ = 20~km/s and $v\sin{i}$ = 25~km/s.  Imprinted $^{13}$CH$_4$ lines on the model are shown as dashed red lines. Several CO features are detected.}
\label{fig:cshell}
\end{figure*}

For Luhman~16AB, high-resolution spectroscopy can constrain association membership (see above), orbital motion and individual component masses, and the orientation of the rotation axis on the sky.  As a first step toward these measurements, we used CSHELL \citep{1993SPIE.1946..313G} on IRTF on 25 April 2013 (UT) to obtain high-resolution
({\ldl} $\approx$ 43,000) NIR spectroscopy over a $\sim$60~{\AA} window centered at 2.3124~{\micron}.  We deployed the $^{13}$CH$_4$ isotopologue gas cell to better determine our wavelength calibration \citep{2012PASP..124..586A}.  The binary was aligned with the 0$\farcs$5 slit, and a sequence of nine 900~s exposures was obtained for both components simultaneously.  A preliminary reduction of the combined light spectrum is shown in Figure~\ref{fig:cshell}. With an average S/N $\approx$ 10, we are able to identify several features associated with CO absorption in the target spectra.  These data can be reproduced with a 1500~K model with $v_{rad}$ $\approx$ 20~km/s (consistent with \citealt{2013ApJ...770..124K}) and $v\sin{i}$ $\approx$ 25~km/s, consistent with the variability period and a nearly equatorial orientation.  

\section{Next Steps}

In addition to NIR spectroscopy, we obtained resolved optical spectroscopy (moderate and high resolution), resolved optical and NIR photometric monitoring, and a deep radio measurement.  Because these observations occurred within a single campaign, we will be able to examine correlations between the measurements (e.g., SEDs at low and high brightnesses, spectral phase variations).  Our detailed focus on Luhman~16AB will hopefully clarify some of the remarkable aspects of the L/T transition.

\begin{acknowledgements}
We are grateful to all of our telescope operators and instrument specialists
for their assistance with the observations, and to A.\ Tokunaga and D.\ Osip 
for allowing observations to be conducted during engineering time. 
\end{acknowledgements}

\bibliographystyle{aa}

\end{document}